\documentstyle[amsfonts,preprint,aps]{revtex}

\begin{document}
\title{Comment on ''Photonic Band Gaps: Noncommuting limit and the 'Acoustic Band'''}
\author{Didier Felbacq}
\address{LASMEA UMR-CNRS 6602\\
Complexe des C\'{e}zeaux\\
63177 Aubi\`{e}re Cedex, France}
\date{\today
}
\maketitle

\tightenlines This comment concerns the homogenization of 2D dielectric
photonic crystals, and the fact that the limits $k\rightarrow 0$ ($k$ is the
Bloch vector) and $\varepsilon \rightarrow +\infty $ do not commute for $p$%
-polarized waves (it is so for $s$-polarized waves, which is a
straightforward case). This result has been claimed to be true by Nicorovici
\& al. in a series of paper \cite{nico1,nico2,nico3} and it has been claimed
to be false by Krokhin \& al. in a comment \cite{kro}. The point of this
note is to make the situation clear one for all, that is to give a
mathematically clean derivation of the result by Nicorovici and to prove
that it is right. The first point is to use a clear definition of a
homogenization process: rather than letting tend the Bloch vector to zero it
is in my opinion clearer to deal with a finite-size photonic crystal,
contained in a bounded domain $\Omega $, with period $\eta $ (the period is
a contracted cell $\eta Y$, where $Y=[0,1[^{2}$ and $\theta $ is the filling
ratio in $Y$ see fig. 1 for notations) and a fixed wavenumber $k_{0}$, in
which case for an incident field $u^{i}$ the total field $u_{\eta }$
satisfies in $p$-polarization div$\left( \varepsilon _{\eta }^{-1}\nabla
u_{\eta }\right) +k_{0}^{2}u_{\eta }=0$, where $\varepsilon _{\eta }$
represents the relative permittivity of the rods which are homogeneous
circular cross-section rods, the permittivity of one rod being equal to $%
\varepsilon _{s}$. Then we study the limit of $u_{\eta }$ when $\eta
\rightarrow 0$ (in case of an infinite crystal with no incident field, this
amounts to let $k$ tend to zero for a Bloch wave). I have shown in a
previous paper \cite{moibou} that $u_{\eta }$ tends to $u_{0}$ satisfying div%
$\left( \varepsilon _{\hom }^{-1}\nabla u_{0}\right) +k_{0}^{2}u_{0}=0$
where 
\begin{equation}
\varepsilon _{\hom }=\left\{ 
\begin{array}{l}
\left( 1+\theta \left( \varepsilon _{s}^{-1}-1\right) +\phi _{\varepsilon
}\right) ^{-1}\text{ in }\Omega  \\ 
1\text{ outside }\Omega 
\end{array}
\right.   \label{hom}
\end{equation}
and $\phi _{\varepsilon }$ is a term defined in \cite{moibou}. If we let
formally tend $\varepsilon _{s}$ to infinity we get $\varepsilon _{\hom
}\rightarrow \left( 1-\theta +\phi _{\infty }\right) ^{-1}$ where $\phi
_{\infty }=\left\langle 
{\displaystyle{\partial w \over \partial y_{1}}}%
\right\rangle _{Y}$, $w$ being the unique $Y$-periodic solution, with null
mean, of the following problem 
\begin{equation}
\left\{ 
\begin{array}{ll}
\Delta w=0 & \text{in }Y\backslash P \\ 
{\displaystyle{\partial w \over \partial n}}%
=-{\bf n}.{\bf e}_{1} & \text{on }\partial P
\end{array}
\right. .  \label{annex}
\end{equation}
\newline
and $\left\langle .\right\rangle _{Y}$ denotes averaging over $Y$.

Now let us deal directly with the infinitely conducting crystal. At step $%
\eta $ the field satisfies $\Delta u_{\eta }+k_{0}^{2}u_{\eta }=0$ in the
complementary of the rods, which are denoted by $T_{\eta }$, and $%
{\displaystyle{\partial u_{\eta } \over \partial n}}%
=0$ on $\partial T_{\eta }$. Our result is

{\bf Theorem}

{\it When }$\eta ${\it \ tends to }$0$,{\it \ }$u_{\eta }${\it \ tends to }$%
u_{0}${\it \ satisfying }div$\left( \varepsilon _{\infty }^{-1}\nabla
u_{0}\right) +k_{0}^{2}u_{0}=0$ {\it where} 
\[
\varepsilon _{\infty }=\left\{ 
\begin{array}{l}
\left( 1-\theta \right) \left( 1-\theta +\phi _{\infty }\right) ^{-1}\text{
in }\Omega \\ 
1\text{ outside }\Omega
\end{array}
\right. 
\]
A simple comparison with (\ref{hom}) leads to the evident conclusion that

{\bf Corollary }

{\it The limits }$\eta \rightarrow 0${\it \ and }$\varepsilon \rightarrow
+\infty ${\it \ do not commute.}

{\bf Proof}

The field $u_{\eta }$ is null inside the rods but we can define a function $%
\widetilde{u}_{\eta }$ such that $u_{\eta }=\widetilde{u}_{\eta }$ outside $%
T_{\eta }$ and $\widetilde{u}_{\eta }$ is in the Sobolev space $%
H_{loc}^{1}\left( {\Bbb R}^{2}\right) $, so that $\left( 1-1_{T_{\eta
}}\right) \widetilde{u}_{\eta }=u_{\eta }$. Using now a test function $\phi $
in the Schwartz space ${\cal D}\left( \Omega \right) $, we find 
\[
-\int_{\Omega }\nabla \widetilde{u}_{\eta }\nabla \phi
\,d^{2}x+k_{0}^{2}\int_{\Omega }\left( 1-1_{T_{\eta }}\right) \widetilde{u}%
_{\eta }\phi \,d^{2}x=0 
\]
\newline
Assuming that $\left( u_{\eta }\right) $ is bounded in $L^{2}\left( \Omega
\right) $ it can easily be shown that $\left( \widetilde{u}_{\eta }\right) $
is bounded in $H^{1}\left( \Omega \right) $ because it satisfies a standard
Helmholtz equation $\Delta \widetilde{u}_{\eta }+k_{0}^{2}\left(
1-1_{T_{\eta }}\right) \widetilde{u}_{\eta }=0$\ . Then, up to the
extraction of a subsequence we have $\widetilde{u}_{\eta }\rightarrow 
\widetilde{u}_{0}$ strongly in $L^{2}\left( \Omega \right) $, $u_{0}=\left(
1-\theta \right) \widetilde{u}_{0}$\ and $\chi _{\eta }=\nabla \widetilde{u}%
_{\eta }\rightharpoonup \chi _{0}$ weakly in $H^{1}\left( \Omega \right) $
so that: $-\int_{\Omega }\chi _{0}\nabla \phi \,d^{2}x+k_{0}^{2}\int_{\Omega
}u_{0}\phi \,d^{2}x=0$ meaning that 
\[
\text{div}\left( \chi _{0}\right) +k_{0}^{2}u_{0}=0. 
\]
We then have to find an expression for $\chi _{0}$. We set $w_{i}=w_{\eta
}+x_{i}$, where $w_{\eta }=w\left( \frac{x}{\eta }\right) $ (note that $w_{i}%
\stackrel{L^{2}}{\rightharpoonup }x_{i}$), we have $-$ $\int_{\Omega }\nabla 
\widetilde{u}_{\eta }\nabla \left( \phi w_{i}\right)
\,d^{2}x+k_{0}^{2}\int_{\Omega }u_{\eta }\phi \,w_{i}\,d^{2}x=-$ $%
\int_{\Omega }\chi _{\eta }\nabla \phi \,w_{i}\,d^{2}x+\int_{\Omega }\nabla
\phi \nabla w_{i}\widetilde{u}_{\eta }~d^{2}x+k_{0}^{2}\int_{\Omega }u_{\eta
}\phi w_{i}~d^{2}x=0$. Then letting $\eta $ tend to $0$, we obtain 
\[
\int_{\Omega }\left[ -\chi _{0}x_{i}+\left\langle \nabla w_{i}\right\rangle
_{Y}.{\bf e}_{i}\widetilde{u}_{0}\right] \nabla \phi
\,d^{2}x+k_{0}^{2}\int_{\Omega }u_{0}\phi \,x_{i}\,d^{2}x=0 
\]
this shows that $\chi _{0}.{\bf e}_{i}=\left( \left\langle \nabla
w\right\rangle _{Y}+\left( 1-\theta \right) {\bf e}_{i}\right) \nabla 
\widetilde{u}_{0}$. The theorem and corollary follow by the rotational
invariance of the problem.

As a conclusion, we might suggest that a good mathematical background should
make it possible to avoid any polemical discussions over these issues.

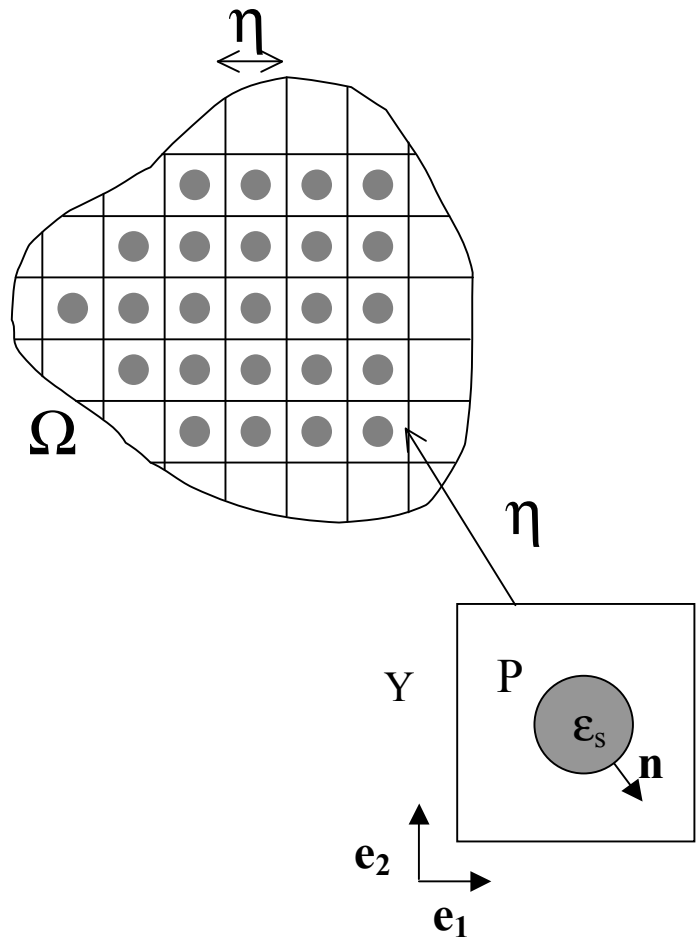
\end{document}